\documentclass[aps,print,twocolumn,showpacs]{revtex4}%
\usepackage{amsfonts}
\usepackage{amsmath}
\usepackage{amssymb}
\usepackage{graphicx}%
\providecommand{\U}[1]{\protect\rule{.1in}{.1in}}
\begin{document}
\preprint{$\pm\div$ }
\title[ ]{Macroscopic Quantum States and Quantum Phase Transition in Dicke Models of
Arbitrary Atom-Number}
\author{Jinling Lian, Yuanwei Zhang, and J.-Q. Liang$^{\ast}$}
\affiliation{Institute of Theoretical Physics and Department of Physics, Shanxi University,
Taiyuan 030006, China}

\begin{abstract}
The energy spectrum of Dicke Hamiltonians with and without the rotating wave
approximation for arbitrary atom-number is obtained analytically with the
variational method, in which the effective pseudo-spin Hamiltonian resulted
from the expectation value in the boson-field coherent state is diagonalized
by the spin-coherent-state transformation. In addition to the ground-state
energy an excited macroscopic quantum-state is found corresponding to the
south-and-north-pole gauges of the spin-coherent states respectively. Our
results of ground-state energies in exact agreement with various approaches
show that these models exhibit a zero-temperature quantum phase transition of
second-order for any number of atoms, which however was commonly considered as
a phenomenon of the thermodynamic limit with the atom-number tending to
infinite. The critical behavior of geometric phase is analyzed, which displays
no singularity at the critical point.

\end{abstract}
\maketitle

In this paper we analyze the energy spectrum of macroscopic quantum-states
(MQS) in relation with the quantum phase transition (QPT) for $N$ two-level
atoms interacting with a single bosonic mode. The system is known as the Dicke
model (DM) \cite{1}, while the MQS refers to boson and spin coherent-states.
Although the model itself is quite simple, it displays a rich variety of the
unique aspects of quantum theory and has become a paradigmatic example of
collective quantum behaviors. Thus DM is of fundamental importance in quantum
optics and any systems involving the interaction between matter and light, for
example, the superconducting circuits coupled to a resonator. Since the atoms
are confined in a small container compared to the wavelength of
radiation-field the model is reduced to a quantum mechanical problem, which
can be solved exactly. At zero temperature, the system of $N$ interacting
atoms via the bosonic field can exhibit a transition to a superradiant phase
called the QPT, which describes a structural change in the properties of the
ground-state energy spectrum \cite{2} at the critical value of coupling
parameter, where both the numbers of field-bosons and atoms in excited states
exhibit an abrupt increase from zero. The QPT has become a fundamental way to
explore dynamic properties of quantum correlations in many-body physics and
thus is of central interest in the realm of quantum information theory
\cite{3} as well. The DM, which exhibits a second-order phase transition from
the normal phase to the superradiant phase predicted long ago \cite{4,5}, has
been studied extensively because of its broad application range such as the
ground-state entanglement and corresponding finite-size behavior. The critical
exponents and finite-size corrections are obtained \cite{6} by a
Holstein-Primakoff (HP) approach \cite{7}, which is the most popular method
\cite{8} in the investigation of QPT of DM.

It is believed that the QPT can take place only if the collective atom-photon
coupling strength is the same order of the atom level-space, which was
considered as a challenging transition-condition. In the strongly coupled
regime of cavity quantum electrodynamics this condition, however, is shown to
be accessible by controlling the pump laser power \cite{9}. For a
Bose-Einstein condensate in a high-finesse optical cavity, the energy space of
two levels can be adjusted to be small enough and the QPT, namely the
superradiance transition, has been observed experimentally \cite{10,16,17}.

The HP representation that converts the Hamiltonian into a two-mode bosonic
problem is the starting point for the most theoretical analysis of the QPT in
the thermodynamic limit ($N\rightarrow\infty$), which leads to solvable
Hamiltonians by neglecting terms from expansions of the HP square roots
\cite{8}. In this paper we present an alternative procedure to study the
energy spectrum of MQSs with the standard variational method. An effective
pseudo-spin Hamiltonian in the energy functional is diagonalized in terms of
the spin coherent state (SCS) transformation \cite{11,12}, which is the key
point to lead to the new observation. Both the energy-spectra of two MQSs
obtained from the south-and-north-pole gauges of SCSs are found to have the
behavior of QPT, where only one is the ground state. In the MQSs the QPT
regarded as a drastic change of ground-state energy spectrum exists for any
number of atoms since the SCS is valid for any spin-value including the
spin-$\frac{1}{2}$ corresponding to the one-atom case and the
thermodynamic-limit approximation resulted from the HP-transformation is
avoided. The critical behaviors of geometric phase (GP) are also analyzed.

\emph{Dicke Hamiltonian and spin-coherent-state representation }Dicke
Hamiltonian (DH) describing the interaction of $N$ two-level atoms of
level-space $\Omega$ with a single bosonic mode of frequency $\omega$, can be
generally written as $H=H_{0}+H_{sb}$, where $H_{0}=\omega a^{\dag}a+\Omega
J_{z},$and the spin-boson coupling part is defined as%
\begin{equation}
H_{sb}=\frac{g}{\sqrt{N}}J_{x}\left(  a^{\dag}+a\right)  .\label{1}%
\end{equation}
The operators $a$ and $a^{\dag}$\ denote the one-mode annihilation and
creation boson (photon)-operators respectively, $J_{z}$ is the atomic
relative-population operator, and $J_{\pm}$ $=(J_{x}\pm iJ_{y})$ are the
atomic transition operators. Various approximations have been developed, which
are suitable for particular ranges of parameters. The rotating-wave
approximation (RWA), which has broad application in quantum optics, is based
on the assumption of near resonance and relatively weak coupling between the
boson-field and atoms \cite{13}. The spin-boson coupling part becomes%
\begin{equation}
H_{sb}^{r}=\frac{g}{2\sqrt{N}}[J_{x}(a+a^{\dag})+iJ_{y}(a-a^{\dag})]\label{2}%
\end{equation}
in the RWA.

We are going to determine the energy-values of the MQSs as a function of the
coupling constant $g$. To this end we use a variational method with the trail
state being a direct product of SCS and Weyl (boson coherent) state \cite{15}
$|\psi_{\pm}\rangle=|\alpha\rangle|\pm\mathbf{n}\rangle$ where the boson
coherent-state is defined by $a|\alpha\rangle=\alpha|\alpha\rangle$ and the
$SU(2)$ SCS is%
\begin{equation}
\mathbf{J}\cdot\mathbf{n}|\pm\mathbf{n}\rangle=\pm s|\pm\mathbf{n}%
\rangle\label{3}%
\end{equation}
with $s=\frac{N}{2}$ being the total pseudo-spin value. The unit vector%
\begin{equation}
\mathbf{n=}(\sin\theta\cos\phi,\sin\theta\sin\phi,\cos\theta)\label{4}%
\end{equation}
expands a Bloch sphere. The states $|\pm\mathbf{n}\rangle$ defined as an
eigenstate of operator $\mathbf{J}\cdot\mathbf{n}$ with eigenvalues $\pm s$
are called the SCSs of north-and-south-pole gauges respectively, in which the
minimum uncertainty $\frac{1}{2}|\langle J_{z}\rangle|=\langle(\Delta
J_{x})^{2}\rangle^{\frac{1}{2}}\langle(\Delta J_{y})^{2}\rangle^{\frac{1}{2}}$
is satisfied, where $\langle J_{z}\rangle=\langle\pm\mathbf{n|}J_{z}%
|\pm\mathbf{n}\rangle$. It has been shown that the expectation value of spin
in the SCS satisfies the exactly same Bloch equation as that of a classical
magnetization-vector in external fields \cite{Liang}, so that we refer it as a
MQS. The SCSs can be generated from the extreme states of $J_{z}$, that
$J_{z}|s,\pm s\rangle=\pm s|s,\pm s\rangle$, by a rotation $|\pm
\mathbf{n}\rangle=R|s,\pm s\rangle$ with the unitary operator obviously being%
\begin{equation}
R=e^{i\theta\mathbf{m}\cdot\mathbf{J}},\label{5}%
\end{equation}
where the unit-vector $\mathbf{m}$ in the $x$-$y$ plane is perpendicular to
the plane expanded by the $z$-axis and unit vector $\mathbf{n}$. The energy
expectation value in the trail state reads%
\begin{equation}
E_{\pm}(\alpha)=\langle\psi_{\pm}|H|\psi_{\pm}\rangle=\langle\pm
\mathbf{n|}H_{es}(\alpha)|\pm\mathbf{n}\rangle.\label{6}%
\end{equation}
The effective spin-Hamiltonian $H_{es}(\alpha)$ resulted from the bosonic
mean-field treatment contains the spin operators only, however, with the
complex bosonic-field parameter $\alpha$ to be determined by the
variation-principle which minimizes the energy functional $E_{\pm}(\alpha)$.
The key point of our method is that we can properly choose the unitary
operator $R$ defined by Eq.(\ref{5}) such that the SCSs $|\pm\mathbf{n}%
\rangle$ become eigenstates of the effective spin-Hamiltonian, i.e.,
$H_{es}(\alpha)|\pm\mathbf{n}\rangle=E_{\pm}(\alpha)|\pm\mathbf{n}\rangle$,
which describes $N$ atoms in an effective, classical light-field parameterized
by $\alpha$. In this stage the energy functional is obtained exactly compared
with the HP approach, where the square root is expanded approximately with
large atom-number $N$. The ground state energy can be obtained by the standard
variational method with respect to the complex parameter $\alpha$,
$\frac{\partial E_{-}(\alpha)}{\partial\alpha}=0$ or $\frac{\partial
E_{+}(\alpha)}{\partial\alpha}=0$, in which only one equation can be satisfied
giving rise to the ground-state energy-spectrum. For the general unit-vector
$\mathbf{n}(\theta,\phi)$ in Eq.(\ref{4}), the vector $\mathbf{m}$ to be
determined is obviously%
\begin{equation}
\mathbf{m=}(\sin\phi,-\cos\phi,0).\label{7}%
\end{equation}

It has been demonstrated by the HP representation that in the thermodynamic
limit ($N\rightarrow\infty$) the system undergoes a QPT at a critical coupling
$g_{c}$, at which point the system changes from a largely unexcited normal
phase to a superradiant one, where all atoms absorb and emit bosons
collectively and coherently \cite{4}. This QPT has recently been observed by
employing a superfluid gas in an optical cavity \cite{10}. The theoretical
description is based on HP realization of the quasi-spin operators by the
infinite expansion-series of additional bosons. The infinite series are
truncated up to the first or second order with the large $N$ assumption, i.e.
the number of excited atoms is small compared with the total number of atoms
\cite{8,9,14}. It is also shown that the number of photons is divergent
\cite{9,14} at the critical value of coupling constant $g_{c}$. We in the
present paper provide the analytical results of the energy spectrum of MQSs,
the occupation-number of atom levels and the expectation value of photon
numbers in terms of the SCS transformation. The critical points of QPT are
found explicitly.

\begin{figure}[h]
\centering \vspace{0cm} \hspace{0cm}
\scalebox{0.5}{\includegraphics{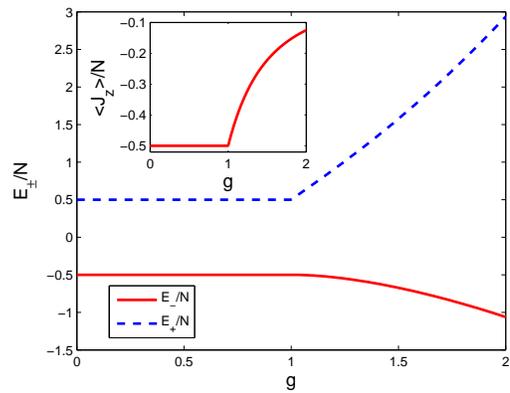}}\caption{(Color online) The
coupling-constant dependence of energy-spectra of MQSs $E_{-}/N$, $E_{+}/N$
(dashed line) and the occupation-probability difference between two
atom-levels $\left\langle J_{z}\right\rangle /N$ (inset) with $\omega
=\Omega=1$ and the critical point $g_{c}=1$.}%
\label{fig1}%
\end{figure}

\emph{Energy spectrum of macroscopic quantum states and quantum phase
transition} For the DM with spin-boson coupling Eq.(\ref{1}), the effective
spin Hamiltonian is%
\begin{equation}
H_{es}(\alpha)=\omega(u^{2}+v^{2})+H_{s}(\alpha),\label{8}%
\end{equation}
where $\alpha=u+iv$, and $H_{s}(\alpha)=\Omega J_{z}+\frac{2gu}{\sqrt{N}}%
J_{x}$ denotes the effective spin Hamiltonian with the constant excluded,
which can be written as $H_{s}(\alpha)=r(\cos\theta J_{z}+\sin\theta J_{x})$
with the$\ $parameters determined as $r=\sqrt{\Omega^{2}+(\frac{2g}{\sqrt{N}%
}u)^{2}}$, $\cos\theta=\frac{\Omega}{r}$, and $\sin\theta=\frac{2gu/\sqrt{N}%
}{r}$. In this case the rotation operator Eq.(\ref{5}) has a simple form with
the unit vector $\mathbf{m}=(0,-1,0)$ seen from Eq.(\ref{7}). The energy
eigenvalue of $H_{s}(\alpha)$ is $E_{s}(\alpha)=sr$ and the total energy
functional is%
\begin{equation}
E_{\pm}(\alpha)=\omega(u^{2}+v^{2})\pm\frac{N}{2}\sqrt{\Omega^{2}+(\frac
{2g}{\sqrt{N}}u)^{2}}.\label{9}%
\end{equation}
The ground-state energy is the minimum of energy functional $E_{-}(\alpha)$
for the SCS of south-pole gauge only. From $\frac{\partial E_{-}(\alpha
)}{\partial v}=0$, $\frac{\partial E_{-}(\alpha)}{\partial u}=0$ we have
$v=0$, and $u=0$ or%
\begin{equation}
u^{2}=|\alpha|^{2}=\frac{N\Omega^{2}}{4g^{2}}\left(  \frac{g^{4}}{g_{c}^{4}%
}-1\right)  ,\label{10}%
\end{equation}
which is photon-number in the superradiant phase. The parameter%
\begin{equation}
g_{c}=\sqrt{\omega\Omega}\label{11}%
\end{equation}
is just the critical point of phase transition and is the same value obtained
with the HB approach \cite{8}. Then the ground-state energy is%
\begin{equation}
E_{-}=\left\{
\begin{array}
[c]{ll}%
-\frac{N\Omega}{2}, & \left(  g\leqslant g_{c}\right)  \\
-\frac{N\Omega g^{2}g_{c}^{2}}{4}\left(  \frac{1}{g_{c}^{4}}+\frac{1}{g^{4}%
}\right)  , & \left(  g>g_{c}\right)  ,
\end{array}
\right.  \label{12}%
\end{equation}
which is also obtained in Refs. \cite{53804,8,15}, however is just an
approximation of thermodynamic limit in the HP-transformation approach. We
demonstrate that the approximation resulted from the HP transformation itself
can be avoided with the SCS representation. Below $g_{c}$ the system is in
normal phase but the superradiation phase above the critical point, while the
QPT emerges for any number of atoms in our formulation and it is no need at
all for the thermodynamic limit. The smooth QPT is of the second order seen
from the continuation of the first-order derivative $\frac{dE_{-}}%
{dg}|_{g=g_{c+}}=\frac{dE_{-}}{dg}|_{g=g_{c-}}=0$ at the critical point (see
Fig. \ref{fig1}). The one half of occupation-number difference between two
levels of the atoms can be evaluated by%
\begin{equation}
\left\langle J_{z}\right\rangle =\langle-s|\widetilde{J}_{z}|-s\rangle
=\left\{
\begin{array}
[c]{ll}%
-\frac{N}{2}, & \left(  g\leqslant g_{c}\right)  \\
-\frac{N}{2}\frac{g_{c}^{2}}{g^{2}}, & \left(  g>g_{c}\right)  ,
\end{array}
\right.  \label{13}%
\end{equation}
where $\widetilde{J}_{z}=R^{\dag}J_{z}R=\cos\theta J_{z}-\sin\theta J_{x}$. It
is interesting to see the energy spectrum of excited MQS resulted from the SCS
of the north-pole gauge that%
\begin{equation}
E_{+}=\left\{
\begin{array}
[c]{ll}%
\frac{N\Omega}{2}, & \left(  g\leqslant g_{c}\right)  \\
\frac{N\Omega g^{2}g_{c}^{2}}{4}\left(  \frac{3}{g_{c}^{4}}-\frac{1}{g^{4}%
}\right)  , & \left(  g>g_{c}\right)  ,
\end{array}
\right.  \label{14}%
\end{equation}
which although cannot be occupied at zero temperature exhibits the behavior of
QPT at the same critical point $g_{c}$. This excited MQS may be of importance
in laser physics and technology. The energy spectra $\frac{E_{\pm}}{N}$ and
the occupation-probability difference $\frac{\left\langle J_{z}\right\rangle
}{N}$ as a function of coupling constant $g$ are plotted in Fig. \ref{fig1}.
For the one-atom case $s=\frac{1}{2}$, $\left\langle J_{z}\right\rangle $ is
nothing but the one half of occupation-probability difference between the two
levels. 

\emph{Rotating wave approximation} In quantum optics the DH is usually studied
in the RWA for small value of coupling, in which the counter-rotating terms
$a^{\dag}J_{+}$ and $aJ_{-}$ are neglected so that the DH becomes integrable.
The effective spin Hamiltonian for the RWA is seen from Eq.(\ref{2}) as%
\begin{equation}
H_{s}^{r}=\Omega J_{z}+\frac{g}{\sqrt{N}}\left(  J_{x}u-J_{y}v\right)  ,
\label{15}%
\end{equation}
which can be rewritten as the standard $\mathbf{J}\cdot\mathbf{n}$ form, i.e.,
$H_{s}^{r}=r\left[  \cos\theta J_{z}+\sin\theta\left(  \cos\phi J_{x}+\sin\phi
J_{y}\right)  \right]  $, where $\cos\phi=\frac{u}{|\alpha|}$, $\sin\phi
=\frac{-v}{|\alpha|}$, $\cos\theta=\frac{\Omega}{r}$, $\sin\theta
=\frac{g|\alpha|}{\sqrt{N}r}$, and $r=\sqrt{\Omega^{2}+g^{2}(u^{2}+v^{2})/N}$.
The unitary operator with the vector $\mathbf{m}$ given in Eq.(\ref{7}) can
transfer the state $|s,\pm s\rangle$ into the eigenstate of $H_{s}$, and the
energy functional in this case becomes%
\begin{equation}
E_{\pm}^{r}(\alpha)=\omega(u^{2}+v^{2})\pm\frac{N}{2}\sqrt{\Omega^{2}%
+\frac{g^{2}(u^{2}+v^{2})}{N}}. \label{16}%
\end{equation}
From $\frac{\partial E_{-}^{r}(\alpha)}{\partial u}=0$, and $\frac{\partial
E_{-}^{r}(\alpha)}{\partial v}=0$ we have $u=v=0$ or%
\begin{equation}
u^{2}+v^{2}=|\alpha|^{2}=\frac{N\Omega^{2}}{g^{2}}\left(  \frac{g^{4}}%
{(g_{c}^{r})^{4}}-1\right)  , \label{17}%
\end{equation}
where the critical point is seen to be%
\begin{equation}
g_{c}^{r}=2\sqrt{\omega\Omega},
\end{equation}
which is two times of the value\ $g_{c}$ in the DH because of neglecting the
counter-rotating terms and is in agreement with the result obtained in Ref.
\cite{5}, where it was found that the phase transition does not exist below
this value. The same critical point is also obtained in Ref. \cite{20} with HP
transformation. The ground-state energy spectrum $E_{-}^{r}$ has the exactly
same form as Eq.(\ref{12}) for the DH, however, with the critical point
replaced by $g_{c}^{r}$ and so does the spectrum of excited MQS $E_{+}^{r}$.
The occupation-number difference $\left\langle J_{z}\right\rangle $ is also
formerly the same as Eq.(\ref{13}).

\begin{figure}[t]
\centering
\vspace{0cm} \hspace{0cm}
\scalebox{0.5}{\includegraphics{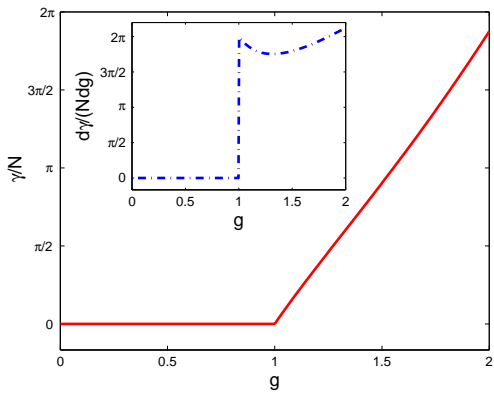}}\caption{(Color online) The
coupling-constant dependence of GP ($\gamma/N$) and its derivative
$d\gamma/(Ndg)$ (inset) with $\omega=\Omega=1$ and the critical point
$g_{c}=1$.}%
\label{fig2}%
\end{figure}

\emph{Critical behavior of geometric phase} Considerable understanding of the
time-evolution of quantum states has been achieved after Berry's discovery of
its geometric feature known as the Berry Phase. An apparently unrelated area
in connection with GP, which has attracted considerable attention recently, is
the study of QPT in many-body systems \cite{18,19,20}. QPT is accompanied by a
drastic change in the ground state at critical points, which is associated
with a non-analytical behavior of the ground-state and is expected to be
reflected in the GP of the wave function. In the DM with various modifications
the GP is generated by a rotation with photon-number operator as in the
quantum optics such that \cite{20,21}%
\begin{align}
\gamma &  =-i\int_{0}^{2\pi}\langle E_{-}|U^{\dag}(\phi)\frac{\partial
}{\partial\phi}U(\phi)|E_{-}\rangle d\phi=2\pi|\alpha|^{2}\nonumber\\
&  =\left\{
\begin{array}
[c]{ll}%
0, & \left(  g\leqslant g_{c}\right)  \\
\frac{\pi N\Omega^{2}}{2g^{2}}\left(  \frac{g^{4}}{g_{c}^{4}}-1\right)  , &
\left(  g>g_{c}\right)  ,
\end{array}
\right.
\end{align}
where the unitary transformation-operator is seen to be $U(\phi)=e^{ia^{\dag
}a\phi}$. The linear $N$-dependence of GP (or photon-number) remains in the
superradiant phase having no particular divergence at the critical point
\cite{15}. The first-order derivative of the GP is obviously%
\begin{equation}
\frac{d\gamma}{dg}=\left\{
\begin{array}
[c]{ll}%
0, & \left(  g\leqslant g_{c}\right)  \\
\pi N\Omega^{2}g\left(  \frac{1}{g_{c}^{4}}+\frac{1}{g^{4}}\right)  , &
\left(  g>g_{c}\right)  .
\end{array}
\right.
\end{equation}
Besides the factor-$2$ difference of critical point ($g_{c}^{r}=2g_{c}$) the
photon number in the RWA case given by Eq.(\ref{17}) has an extra factor-$4$
compared with the photon-number formula Eq.(\ref{10}) for the DH. Fig.
\ref{fig2} shows the GP ($\gamma/N$) and its derivative ($\frac{d\gamma}{Ndg}%
$) (inset) as functions of $g$ for the DH. Unlike the energy spectrum, the GP
exhibits a sharp change at the critical point with discontinuous first-order
derivative. The scaling behavior of the GP at the critical point can be
found\ as%
\begin{equation}
\frac{\gamma}{N}|_{g\longrightarrow g_{c}}=\frac{2\pi\Omega^{2}}{g_{c}^{3}%
}|g-g_{c}|,
\end{equation}
which agrees with the result of HP representation in the thermodynamic limit
\cite{20} and shows no singularity at the critical point confirming the
conclusion in Ref. \cite{15}. 

\emph{Conclusion} The MQSs are useful in the description of QPT, which is, as
a matter of fact, a macroscopic quantum phenomenon. The effective pseudo-spin
Hamiltonian can be diagonalized by the SCS transformation and the energy
spectra of ground-state along with an excited MQS for any number of atoms are
obtained, which possess the exactly same form of formulas for both DHs with
and without the RWA expressed in terms of the critical-point parameter, while
the critical points themselves have a factor-$2$ difference. The SCS technique
universal for various Hamiltonians of Dicke-type, is valid for strong
spin-boson coupling while has no restriction on the number of atoms. Thus we
conclude that the abrupt change of ground-state energy spectrum regarded as
QPT is an effect of strong light-field, which is independent of the atom
number in DHs. At the critical point of phase transition, the expectation
values of photon and atom numbers in excited states exhibit an drastic
increase from zero proportional to the total number of atoms $N$ and remain
the same in the superradiant phase. There is no singular behavior at the
critical point as demonstrated in Ref. \cite{15}, where the tensorial product
of spin-coherent and Weyl states firstly used to analyze the ground-state
energy of DM is shown to lead to an energy surface of the DH with four
variables and two parameters. We adopt the SCS transformation to obtain the
analytic energy spectra of ground state and an excited MQS as well for both
the Hamiltonians with and without the RWA.

\emph{Acknowledgment} This work was supported by National Nature Science
Foundation of China (Grant No.11075099).

\bigskip*jqliang@sxu.edu.cn

\end{document}